# Experimental Investigation of the Photochemical Production of Hydrocarbons in Warm Gas Giant Exoplanet Atmospheres

Benjamin Fleury[1,2,*], Yves Benilan[2], Olivia Venot[3], Bryana L. Henderson[1], Mark Swain[1], Murthy S. Gudipati[1*]

[1] Jet Propulsion Laboratory, California Institute of Technology, 4800 Oak Grove Drive, Pasadena CA 91109

[2] Univ Paris Est Creteil and Université Paris Cité, CNRS, LISA, F-94010 Créteil, France

[3] Université Paris Cité and Univ Paris Est Creteil, CNRS, LISA, F-75013 Paris, France

*Corresponding authors: benjamin.fleury@lisa.ipsl.fr ; murthy.gudipati@jpl.nasa.gov

**Abstract**

In warm (equilibrium temperature < 1000 K), gas giant exoplanet atmospheres, the observation of trace species in abundances deviating from thermochemical equilibrium predictions could be used as an indicator of disequilibrium chemical processes such as photochemistry. To predict which compounds could be used as such tracers, it is therefore essential to study how photochemical processes affect their abundances. For this purpose, we investigated experimentally the efficiency of the photochemical formation of hydrocarbons in gas mixtures representative of warm gas giant atmospheres as a function of the gas temperature at millibar pressures. We find that, compared to thermal reactions alone, photochemistry efficiently promotes, under the studied conditions, the formation of hydrocarbons, with the detection of acetylene, ethane, and propane, as well as carbon monoxide. Therefore, our results confirm the importance of photochemistry in exoplanet atmospheres as a disequilibrium process. Ethane is the major hydrocarbon formed in our experiments, in apparent contradiction with the prediction by thermo-photochemical models that acetylene should be the main hydrocarbon product. We also observe an evolution of the hydrocarbon production efficiency as a function of the temperature, a behavior not reproduced by a 0D thermo-photochemical model. Additional studies are necessary to definitively understand the origin of the differences between experimental and modeling results and infer the importance of our results for understanding hydrocarbons formation in warm gas giant exoplanet atmospheres. Finally, our work demonstrates the importance of experimental studies together with modeling studies to accurately interpret, understand, and predict observations of exoplanet atmospheres.

## 1. Introduction

Based on more than two decades of observations and theoretical studies, gas giant exoplanets (i.e., Neptune- to Jupiter-sized planets) are considered to have dense atmospheres mainly constituted of molecular hydrogen ($H_2$) and helium (He). This atmospheric composition has been inferred based on the mass-radius relationship of observed planets. Indeed, for a given mass, if a planet exhibits a radius larger than the one of a pure water world, it implies that a substantial fraction of this planet is made of lower-density constituents, of which $H_2$ and He are the most abundant (Leconte et al. 2015). This is also supported by the observation of atomic hydrogen (H) (Lecavelier des Etangs et al. 2010; Vidal-Madjar et al. 2003; Yan & Henning 2018) and neutral He (Allart et al. 2018; Nortmann et al. 2018; Spake et al. 2018) escaping from the atmospheres of some of these planets orbiting very closely to their host stars and receiving high X-ray and UV fluxes that can cause hydrodynamical escape (Owen 2019). In addition, these atmospheres may contain various trace gas species including methane ($CH_4$), carbon monoxide (CO), water vapor ($H_2O$), or carbon dioxide ($CO_2$), as suggested by various observations (Giacobbe et al. 2021; Guilluy et al. 2019; Swain et al. 2009a; Swain et al. 2009b; Tinetti et al. 2007). Recently, $H_2O$, CO, $CO_2$, and sulfur dioxide ($SO_2$) have been detected by JWST's observations as part of the Early Science Release program (Ahrer et al. 2023; Alderson et al. 2023; Rustamkulov et al. 2023; Tsai et al. 2023). Assuming chemical equilibrium, the abundances of these trace gas species could be theoretically predicted by atmospheric models (Burrows & Sharp 1999; Lodders & Fegley Jr 2002), based on constraints on the temperature, pressure, and bulk composition. However, disequilibrium chemical processes such as photochemistry may affect the atmospheric composition by dissociating and ionizing the main constituents of the atmosphere, particularly in the upper atmospheric layers (e.g., thermosphere) where the pressure is low and UV photons can penetrate significantly (Madhusudhan et al. 2016; Moses 2014).

Whether gas giant exoplanet atmospheres may depart significantly from thermal equilibrium composition has been investigated by various theoretical studies (Baeyens et al. 2022; Line et al. 2011; Line et al. 2010; Molaverdikhani et al. 2019; Moses et al. 2013b; Moses et al. 2016; Moses et al. 2011; Steinrueck et al. 2019; Venot et al. 2015; Venot et al. 2012). These studies suggest that photochemistry could potentially alter the chemical composition of these atmospheres: on the one hand by destroying major molecules such as CO, $H_2O$, or $CH_4$ and on the other hand, by enhancing the formation of more complex species such as acetylene ($C_2H_2$), hydrogen cyanide (HCN), and nitriles or heavier hydrocarbons such as benzene ($C_6H_6$) (Moses, et al. 2013b; Moses, et al. 2016; Venot, et al. 2015; Venot et al. 2020b).

These disequilibrium processes have been considered when analyzing some observational data, highlighting that in the case of highly irradiated exoplanets, photochemistry may be responsible for an observed chemical composition departing from the one predicted by thermochemical models (Changeat et al. 2022; Knutson et al. 2012; MacDonald & Madhusudhan 2017; Roudier et al. 2021; Stevenson et al. 2010; Swain et al. 2010). First, disequilibrium chemistry such as the photochemical destruction of $CH_4$ has been proposed to explain methane's low abundance in the atmosphere of some exoplanets such as GJ 436b (Madhusudhan & Seager 2011; Stevenson, et al. 2010), a hot-Neptune-like exoplanet with a ~700 K effective temperature. At this temperature, thermochemical models predict $CH_4$ to be the main carbon carrier, in contradiction with observations that suggest an atmosphere depleted in $CH_4$ with a higher abundance of CO. However, this photochemical destruction hypothesis has remained controversial, as photochemical models have failed to reproduce the methane destruction rates required to reproduce the observed abundance of $CH_4$ (Line, et al. 2011; Moses et al. 2013a). More recently, the observation by JWST of sulfur dioxide ($SO_2$) in the atmosphere of WASP-39 b has been interpreted as the first direct evidence of photochemistry happening in an exoplanet atmosphere (Tsai, et al. 2023).

In addition, numerous observations suggest that aerosols are ubiquitous in a large variety of exoplanet atmospheres (Knutson et al. 2014; Kreidberg et al. 2014; Sing et al. 2016), including gas giant exoplanets, and may exhibit a temperature dependence in some cases. Indeed, on the one hand, the recent studies of Estrela et al. (2022) and Crossfield and Kreidberg (2017) have suggested that temperatures below 1000 K could be correlated with more optically thick, photochemically produced hazes. But, on the other hand, Dymont et al. (2022) analyzed the transmission spectra of 25 warm exoplanets and did not find a statistically significant correlation between the haziness of the planets and any planetary or stellar parameters, including the equilibrium temperature. Estrela, et al. (2022) also find that the aerosols are broadly distributed through the atmospheric column. This is consistent with previous works based on a microphysics model, which find that photochemical aerosol formation peaks at around microbar pressures and that sedimentation proceeds to distribute the aerosols through the atmospheric column (Lavvas & Koskinen 2017; Lavvas et al. 2019). Therefore, the types of aerosols (i.e., condensate clouds or photochemical hazes) and their properties remained unconstrained by these observations.

It is therefore clear that the physical processes controlling atmospheric composition and haze/aerosol production in these exoplanet atmospheres remain largely unconstrained and the available data remain too limited to support conclusive results on how the atmospheric thermo- and photochemistry correlates to other planetary and atmospheric parameters (Roudier, et al. 2021; Tsiaras et al. 2018). To bridge this gap, laboratory experiments are an important complementary tool that can advance our understanding of the photochemical processes and aerosol properties in exoplanet atmospheres. Recently, several experimental studies have focused on measuring the formation efficiency and properties of aerosol particles produced in a variety of conditions from cool terrestrial planets to hot gas giant planets (Fleury et al. 2019, 2020; Gavilan et al. 2017; Gavilan et al. 2018; He et al. 2018a; He et al. 2018b; He et al. 2020a; He et al. 2020b; Hörst et al. 2018; Moran et al. 2020; Vuitton et al. 2021) while other studies have focused on ion chemistry in super-Earth-like and sub-Neptune-like exoplanets (Bourgalais et al. 2020; Bourgalais et al. 2021).

In our previous studies, we have investigated experimentally the influence of photochemistry on the composition and the formation of aerosols in hot gas giant exoplanet atmospheres with $T > 1000$ K and different C/O ratios (Fleury, et al. 2019, 2020). In the present study, we extend our experiments to reproduce cooler atmospheres ($T < 1000$ K) for which $CH_4$ is expected to be the main carbon carrier (Lodders & Fegley Jr 2002; Venot, et al. 2015) instead of CO that was the main carbon carrier for the higher temperatures that we investigated previously. This key difference could lead to the more efficient production of hydrocarbons such as acetylene ($C_2H_2$) or ethane ($C_2H_6$) from $CH_4$ photochemistry (Line, et al. 2011). Warmer (500 K < $T$ < 1000 K) and oxygen-poor atmospheres make these exoplanets good candidates for the detection of tracers of atmospheric photochemistry (Venot, et al. 2015). Hence, we present here a detailed investigation of the dependence of atmospheric temperature on hydrocarbon production in warm gas giant exoplanet atmospheres. The manuscript is organized as follows: in Section 2, we present the experimental methodology used in this study. In Section 3 we show the evolution of hydrocarbon production as a function of gas temperature. In Section 4, we discuss the implications of our findings for our understanding of chemistry in warm gas giant atmospheres and, finally, we present our conclusions.

## 2. Methods

### 2.1. UV Irradiation of Gas Mixtures at High Temperatures

To study experimentally the photochemistry in warm gas giant planet atmospheres, we used the Cell for Atmospheric and Aerosol Photochemistry Simulations of Exoplanets (CAAPSE), which has been described in detail in Fleury, et al. (2019). The setup consists of a 48 cm long reaction cell constituted by a quartz tube that is closed at each extremity by stainless steel flanges equipped with $MgF_2$ windows. The cell is enclosed in a tube furnace that can warm up the cell and its contents from ambient temperature (~295 K) to 1873 K. Before each experiment, the cell was pumped and degassed at 100 K higher than the maximum working temperature applied in this study, of 1073 K, by heating to and holding at 1173 K for 24 hr. After this operation, the background pressure at ambient temperature was ~ $2 \times 10^{-8}$ mbar. The purpose of this procedure (i.e., pumping and heating of the cell) at a higher temperature than experimental temperature is to remove any adsorbed gases and purge the cell from atmospheric constituents (e.g., $H_2O$ or $O_2$) that could be present in the cell and thus minimizing the risk of contamination of the gas mixtures used for our experiments.

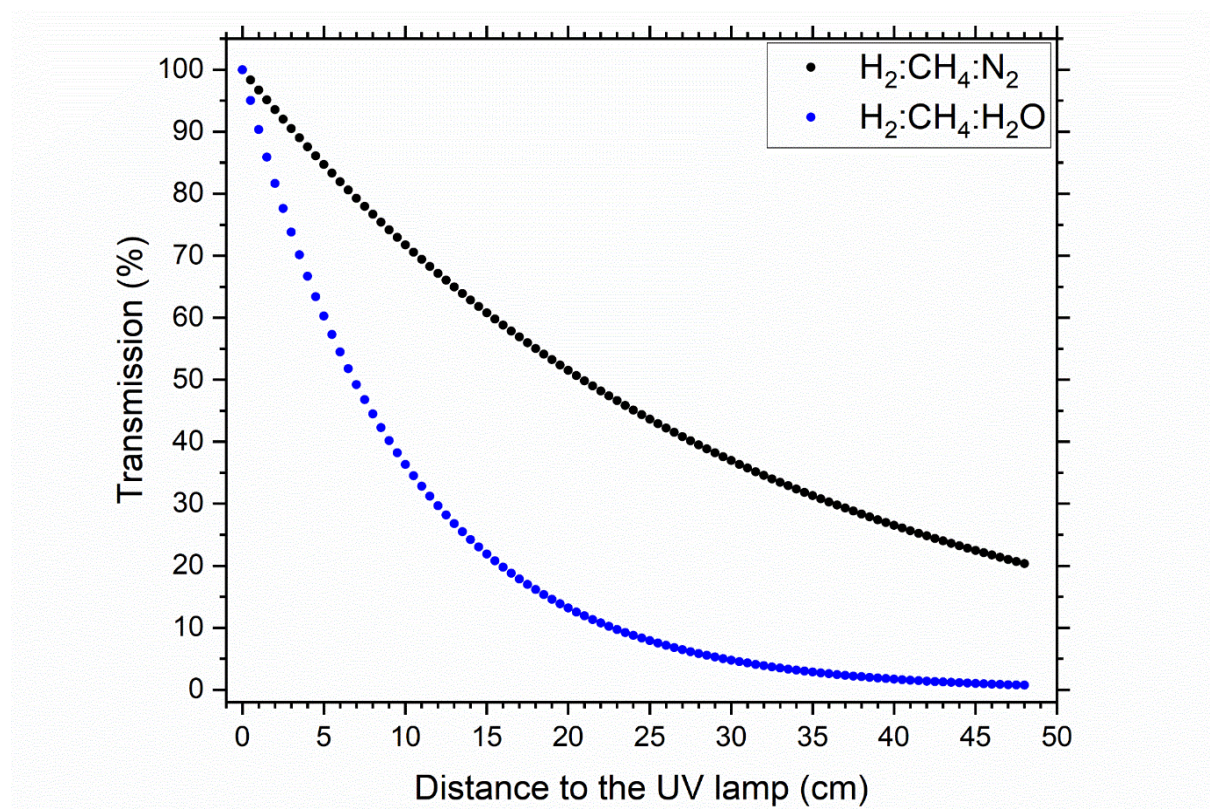

*Figure 1: Transmission of UV photons at $Ly_\alpha$ (121.6 nm) through the 48 cm long gas cell as a function of the distance to the UV lamp at the beginning of the experiments, considering the absorption by $CH_4$ for the $H_2$:$CH_4$:$N_2$ gas mixture and by $CH_4$ and $H_2O$ for the $H_2$:$CH_4$:$H_2O$ gas mixture.*

In this work, we used two gas mixtures to study the effect of the gas composition on the photochemistry. The first gas mixture was made of $H_2$ (Cambridge Isotope Laboratories, 99.99%), $N_2$ (Cambridge Isotope Laboratories, 99.95%), and $CH_4$ (Cambridge Isotope Laboratories, 99.5%), with mixing ratios by volume of 99%, 0.5%, and 0.5%, respectively. The second mixture was made of $H_2$ (Cambridge Isotope Laboratories, 99.99%), $H_2O$ (Cambridge Isotope Laboratories, 99.95%), and $CH_4$ (Cambridge Isotope Laboratories, 99.5%), with mixing ratios by volume of 98.4%, 0.8%, and 0.8%, respectively. These compositions were chosen based on the main atmospheric constituents predicted in Venot, et al. (2015) using a thermo-photochemical model applied to an atmosphere with a temperature of 500 K in the photodissociation region and a C/O ratio of 0.54. This corresponds to a commonly used solar value of the C/O ratio, although some disagreement exists on the exact value (Asplund et al. 2009; Caffau et al. 2011; Lodders 2003, 2010). Although under thermal equilibrium, the gas mixing ratios would change when the temperature increases, we decided to keep the same gas mixture composition at every temperature to measure the effect of the temperature on the chemistry of our specific gas mixture composition. We decided to perform two sets of experiments with gas mixtures containing or not containing $H_2O$, to infer its role in the chemistry of these warm gas giant atmospheres. However, we encountered experimental difficulties when performing the irradiation experiments with the $H_2$:$CH_4$:$H_2O$ mixture, leading to inconclusive results. These difficulties will be described in more detail in Section 3.1. As in Fleury, et al. (2019, 2020), the gases were premixed in a 2 l glass bulb and introduced into the cell at ambient temperature until the pressure in the cell reached 15 mbar (16 mbar in the case of the $H_2$:$CH_4$:$N_2$ experiment at 473 K).

After the introduction of the gas at room temperature, the cell was heated at a rate of 5 K min$^{-1}$ to the different studied temperatures, 473, 673, 873, and 1073 K and then the temperature was held for ~22 hr to ensure that a thermal equilibrium was reached. As in Fleury, et al. (2019, 2020), we observed an increase of the pressure in the cell after heating the gas. The total pressures measured in the cell after the 22 hr of heating are reported in Table 1. In these conditions, we do not expect that the observed increase of the pressure reveals any change in the gas-phase composition.

*Table 1: Total pressure ($P_{heat}$) measured in the reaction cell after the introduction for each experiment of 15 mbar of gas mixture at ambient temperature (16 mbar for the experiment done*

*with $H_2$:$CH_4$:$N_2$ at 473 K) followed by 22 hr at ambient temperature and at the different set temperatures 473, 673, 873, and 1073 K.*

|  | $H_2$:$CH_4$:$N_2$ | $H_2$:$CH_4$:$H_2O$ |
|---|---|---|
| $T$ (K) | $P_{heat}$ (mbar) | $P_{heat}$ (mbar) |
| 295 | 15 | 15 |
| 473 | 19 | 19 |
| 673 | 22 | 22 |
| 873 | 26 | 24 |
| 1073 | 29 | 29 |

Thereafter, we irradiated the heated gas mixture for 24 hr using a microwave-discharge hydrogen-flow lamp. This type of lamp produces UV radiation dominated by Ly$_\alpha$ at 121.6 nm and $H_2$ molecular emission in the 145-165 nm regions (Chen et al. 2014; Es-sebbar et al. 2015; Ligterink et al. 2015). Thus, the photons emitted by the lamp have enough energy to directly photodissociate $H_2O$ and $CH_4$ but not $N_2$ and $H_2$. However, some experimental studies have observed, with similar sources of photons, the activation of $N_2$ chemistry via an unknown mechanism (Trainer et al. 2012; Yoon et al. 2014) at higher pressures (173 mbar to near-atmospheric pressures) than were used here (14 to 29 mbar).

As described in detail in Fleury, et al. (2019), a temperature gradient exists along the cell. The temperature profiles obtained in Fleury, et al. (2019) for oven temperatures set to 573, 873, and 1173 K are presented with more details in Section 2.2. The temperature is at a maximum (and equal to the set oven temperature) at the center of the cell, and then it decreases toward the extremities of the cell, which are cooled to the ambient temperature (~295 K). This agrees with the pressure increase observed in the cell, which is not proportional to the maximum temperature in the cell but rather to a mean temperature of the gas reflecting the existence of this temperature gradient. In our initial gas mixtures, $CH_4$ and $H_2O$ are the dominant absorbers, while $H_2$ and $N_2$ do not absorb in the wavelength range of the irradiation. $CH_4$ and $H_2O$ have strong absorptions at Ly$_\alpha$ and lower absorptions in the 155-165 nm region (Chen & Wu 2004; Mota et al. 2005), where photons are also emitted by the lamp. We calculated the percentage of transmission of Ly$_\alpha$ UV photons, which are the most absorbed by the gases, through the gas cell at the beginning of the experiments as a function of the distance to the UV lamp for the two gas mixtures, considering the absorption by $CH_4$ only for the $H_2$:$CH_4$:$N_2$ gas mixture and by $CH_4$ and $H_2O$ for the $H_2$:$CH_4$:$H_2O$ gas mixture. For the calculation, we used a density at ambient temperature (determined with IR spectroscopy, see Table 3) for $CH_4$ of $1.8 \times 10^{15}$ molecule cm$^-$

³ in the case of the $H_2$:$CH_4$:$N_2$ gas mixture and densities for $CH_4$ and $H_2O$ of $3.09 \times 10^{15}$ molecule cm$^{-3}$ each in the case of the $H_2$:$CH_4$:$H_2O$ gas mixture. We used absorption cross-sections determined at 121.6 nm and ambient temperature of ∼$1.8 \times 10^{-17}$ cm$^2$ for $CH_4$ (Chen & Wu 2004) and $1.47 \times 10^{-17}$ cm$^2$ for $H_2O$ (Mota, et al. 2005). The results are presented in Figure 1. In the case of the $H_2$:$CH_4$:$N_2$ gas mixture, we can see that ∼45% of Ly$_\alpha$ photons reach the center of the cell, where the molecules are at the highest temperature, and about ~20% of the photons go through the entire cell. Because of the addition of $H_2O$, which also absorbs at Ly$_\alpha$, in the case of the $H_2$:$CH_4$:$H_2O$ gas mixture, the transmission decreases to ~9% at the center of the cell and only ~1% of the photons reach the other side of the cell. This rough calculation confirms that UV photons emitted by the lamp are not totally absorbed and can reach the center of the cell and irradiate the gas at higher temperature. Because the conditions in this chamber include a gradient of temperatures from ambient temperature up to the set point, they do not represent any one specific temperature in an exoplanet atmosphere. Nevertheless, these experiments do allow us to study the effect of increasing temperature on atmospheric chemistry, which is the main goal of this study.

### 2.2. Gas-Phase Composition Analysis by IR Spectroscopy

A Thermo Scientific Nicolet iG50 FTIR spectrometer was used, with IR spectroscopy in transmission, to analyze the composition of the gas phase at the different stages of the experiment: after filling the cell with gas mixture at ambient temperature, after 22 hr of heating, and after 24 hr of subsequent irradiation. In our experimental configuration, the IR beam exited the spectrometer, passed through the cell, and was finally detected by a Mercury-Cadmium-Telluride detector cooled to 77 K with liquid nitrogen. For each single spectrum recorded, the absorbance was calculated using as reference a blank spectrum acquired at the beginning of the experiments while the cell was at ambient temperature and under vacuum.

To quantify the molecular abundances of each species, we simulated the IR absorption of each molecule individually along the pathlength. To consider the temperature gradient inside the cell, the 48 cm pathlength was divided into 1 cm segments each having its own temperature. The temperature profiles along the cell were derived from the discrete temperature measurements made at different positions along the quartz tube for set oven temperatures of 573, 873, and 1173 K using the thermocouples in Fleury, et al. (2019). From these data points, the temperature of the cell $T_{cell}(x)$ is modeled using Eq.1. from (Venot et al. 2013a):

$$T_{cell}(x) \; = \; T_{max} + \frac{(T_{amb} - T_{max})}{1 + exp\left(-\frac{|x| - |x_0|}{\Delta x}\right)} \qquad (1)$$

where $T_{amb}$ is the ambient temperature (295 K); $T_{max}$ (K), $x_0$ (cm), and $\Delta x$ (cm) are determined by minimizing the $\chi^2$ function using the measured data. Temperature profiles obtained for these three temperatures are shown in Figure 2. Finally, temperature profiles for the other set oven temperatures used in this study were interpolated from these three temperature profiles.

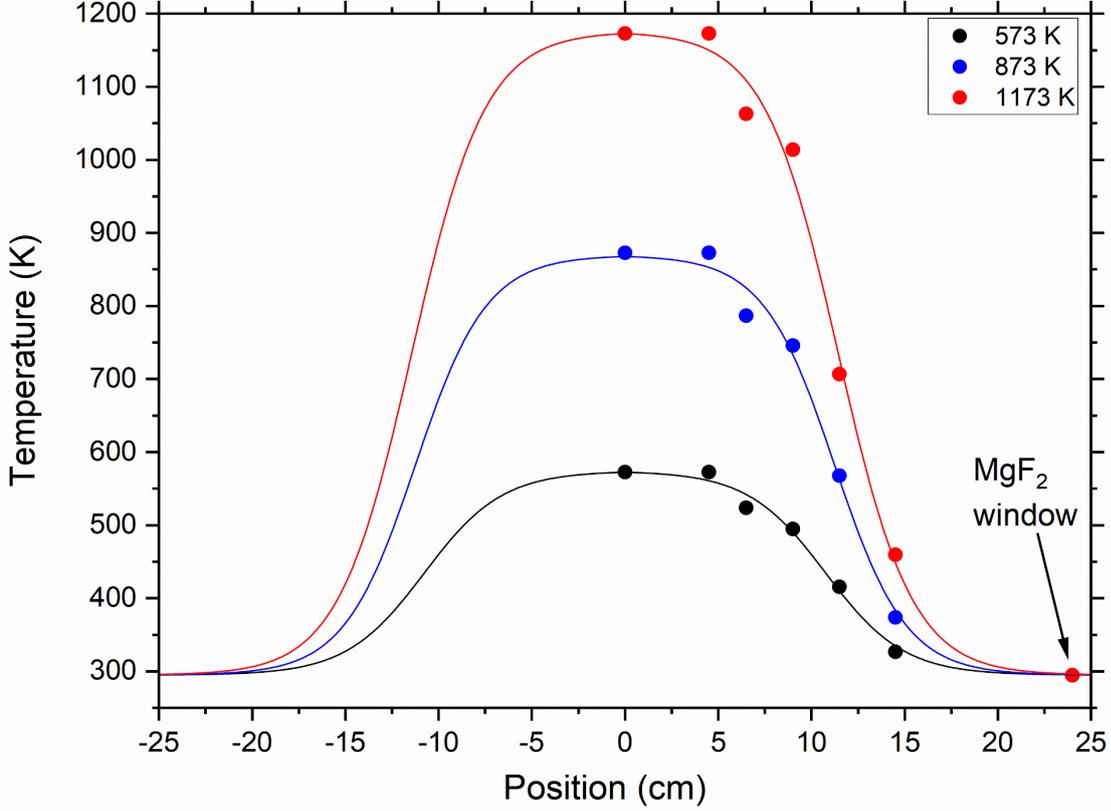

*Figure 2: Temperature measured in Fleury, et al. (2019) at different positions of the quartz tube for different set oven temperature: 573, 873, and 1173 K. Temperature profiles were obtained using Eq. 1. 0 cm corresponds to the center of the cell, where the temperature is maximal ($T_{max}$).*

We assumed that the pressure was constant along the cell, and we then calculated the molecular density in each 1 cm segment based on the pressure and the temperature obtained from the temperature gradient. The molecular data (line positions, line intensities, and air-broadened half-width at half-maximum) were taken from the HITRAN2020 database (Gordon et al. 2022) for $CH_4$, $C_2H_6$, $C_2H_2$ and CO. Each molecular line is simulated using a Voigt profile with a typical sampling of a few $10^{-4}$ cm$^{-1}$. All lines are then summed up to calculate the absorption at a given wavenumber and a given temperature in a 1 cm slice. All the absorptions are then added up to simulate the transmission through the entire cell. We then convolved this transmission with the instrumental function to compare with the experimental data. The comparison is made through a Levenberg-Marquardt algorithm. Because all the absorptions are added up, this implies that the final simulated absorbance spectrum has contributions from both the absorption by the cold gas at the extremities of the cell and the hot gas at the center of the cell. We decided to evaluate the contribution of the gas from different domains of temperature along the optical path to the total absorption of the gas along the entire cell. We extracted the absorbance of the gas calculated for each segment of 1 cm during the simulation of the spectrum of $CH_4$ for the $H_2$:$CH_4$:$N_2$ (99%:0.5%:0.5%) gas mixture after 22 hr at a set temperature of 673 K. The segments of 1 cm were grouped into four temperature regions, 295-373, 374-473, 474-573, and 574-673 K and the absorbance from the segments inside a region summed. Finally, the absorbance of each region was divided by the total absorbance of the simulated spectrum to obtain the contributions (in percent) from the different temperature regions of the cell to the simulated spectrum. The results are presented in Figure 3.

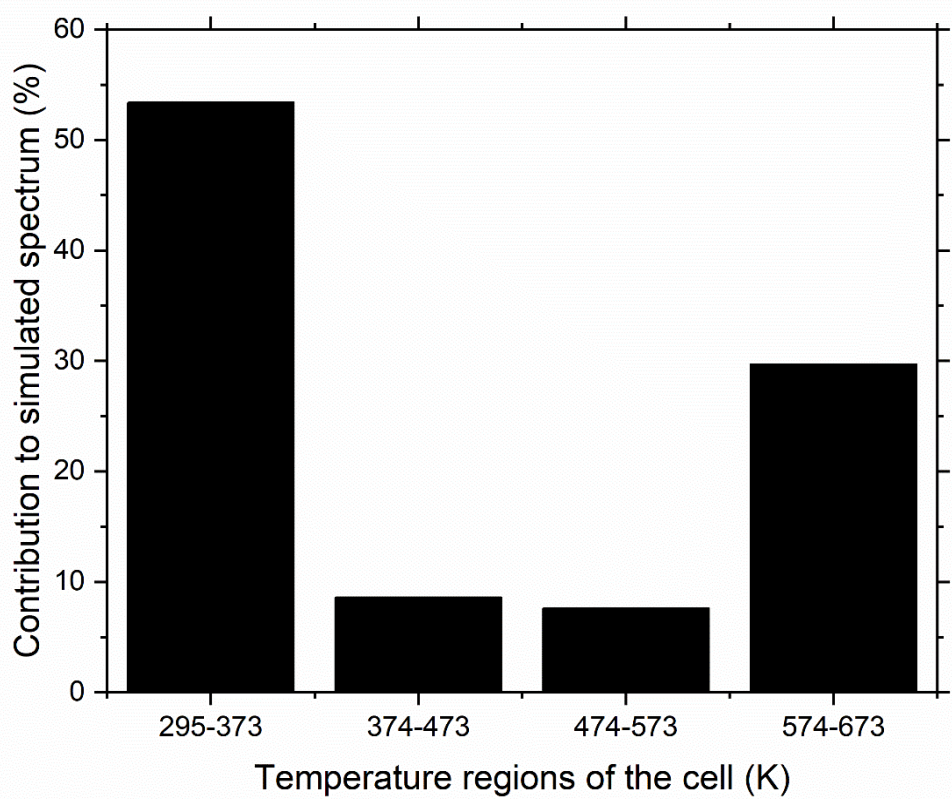

*Figure 3 : Contributions from the different temperature regions of the cell to the spectrum of $CH_4$ simulated for the $H_2$:$CH_4$:$N_2$ (99%:0.5%:0.5%) gas mixture after 22 hr at a set temperature of 673 K. The simulated spectrum is shown in Figure 8.*

We observed that the cold gas at the extremities of the cell (295 to 373 K) contributes 53% of the total absorbance of the gas inside the cell, while the hot gas in the center (574 to 673 K) contributes 30%. The rest of the contribution (27%) comes from the gas at intermediate temperatures. Because we assumed that the pressure is constant along the cell, the density of the gas varies with the temperature and is higher in the cold part of the cell, explaining why the cold gas is such a large contributor to the absorbance of the gas along the optical path.

### 2.3. 0D Photochemical model

Complementarily, we have run a series of numerical simulations using a 0D thermo-photochemical model with parameters (gas composition, temperature, etc.) mimicking our experimental conditions to support the interpretation of the experimental results. We used the 0D version of a kinetic model developed for hot exoplanet atmospheres with the most updated chemical scheme (Venot et al. 2020a). The chemical scheme contains 108 species, involved in 1906 reactions. In the absence of disequilibrium processes, thermochemical equilibrium is reproduced with kinetics. In addition to these reactions, 55 photodissociation reactions are included. In this 0D model, a cell of gas at fixed pressure and temperature was simulated. We ran two series of simulations: one starting with an initial mixture of gas similar to the initial experiments ($H_2$ at 99%, $CH_4$ at 0.5%, and $N_2$ at 0.5%) and another one also including $H_2O$ with an initial mixing ratio of $10^{-4}$ to investigate the effect of $H_2O$ possibly contaminating the experimental gas mixture (see Section 3) on the chemistry. In all the simulations, the pressure was fixed at 15 mbar, but the temperature of the simulations varied between 295, 473, 673, 873, and 1073 K. The irradiation was modeled with the spectrum of an $H_2$ lamp from Ligterink, et al. (2015), which is a good proxy for the lamp used in the experiment. The flux was scaled so that the model matched the order of magnitude of the methane consumption efficiency measured experimentally after 24 hr of irradiation (Table 3) at 295 K. We ran our models and followed the temporal evolution of the chemical composition for more than $10^5$ s (to simulate the 24 hr of experiments). In the models, we decided to track the evolution of the initial reactant $CH_4$, hydrocarbons with two carbon atoms (i.e., $C_2H_2$ and $C_2H_6$), and carbon monoxide.

## 3. Results

### 3.1. Experimental Gas-phase Composition at Thermal Equilibrium

We monitored the evolution of the gas-phase composition after the heating of the gaseous mixtures using absorption IR spectroscopy of the entire cell.

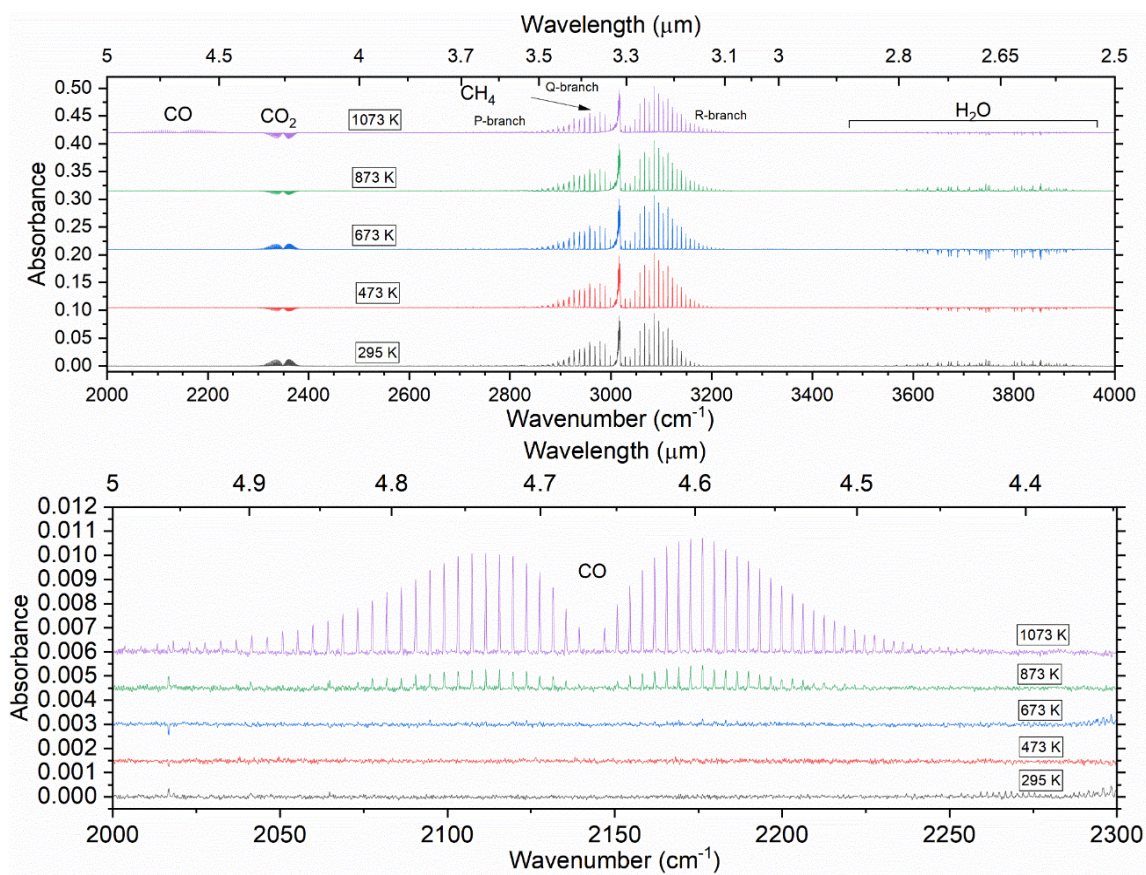

*Figure 4: IR absorbance spectra of the initial gas mixture $H_2:CH_4:N_2$ (99%:0.5%:0.5%) at ambient temperature (black) and after 22 hr of heating at different set temperatures: 473 K (red), 673 K (blue), 873 K (green), and 1073 K (violet). Spectra have been offset for clarity.*

Figure 4 presents the IR absorbance spectra in the 2000-4000 cm$^{-1}$ (2.5–5.0 µm) range of the initial gas mixture $H_2:CH_4:N_2$ at ambient temperature (295 K) and after 22 hr of heating at different set temperatures: 473, 673, 873, and 1073 K. In the initial spectrum at ambient temperature, we observe a band system centered at 3017 cm$^{-1}$ (3.314 µm). This system is composed of a central peak corresponding to the Q branch of the $\nu_3$ C-H stretching band of $CH_4$. On each side of this peak, several rotational transitions of the P and R branches of this same $\nu_3$ band are also visible. At higher temperature, we observe a decrease of the Q branch intensity and a broadening of the peak. In addition, we observed for the P and R branches that higher rotational states are thermally populated at higher temperatures, demonstrating that the gas in the cell is heated. $H_2$ and $N_2$ are inactive in the IR domain, so we do not observe these species in the IR spectra. In addition to $CH_4$, we observed two band systems centered at 2348 cm$^{-1}$ (4.258 µm) and 3785 cm$^{-1}$ (2.642 µm). These absorption bands can be attributed to variations of $CO_2$ and $H_2O$ abundances that are present in the air on the optical pathways, outside of the cell (atmospheric). For the three lowest studied temperatures (i.e., 295, 473, and 673 K), neither the consumption of $CH_4$ nor the formation of products were observed, demonstrating that no thermochemistry occurred. However, for the experiments done at 873 K and 1073 K, we observed the apparition of a new absorption band centered at 2143 cm$^{-1}$ (4.666 µm) that can be attributed to the $\nu_3$ stretching band of CO. The most likely source of CO in our experiments is an oxidation of $CH_4$ by $H_2O$ and possibly enhanced by surface reactions. $H_2O$ molecules could have been adsorbed on the walls of the extremities of the cell and not eliminated during the degassing of the CAAPSE setup at 1173 K prior to the experiments (see Section 2.1) because the extremities of the cell remain colder than the center. Alternatively, a limited amount of $H_2O$ and/or $O_2$ could be slowly leaking with air into the chamber during the duration of the experiment. To determine if the CO production results from methane oxidation, we repeated in the same conditions (duration, pressure, etc.) the experiment at 1073 K, the temperature for which the highest formation of CO has been observed, with a gas mixture made of $H_2$ and $^{13}CH_4$ instead of $^{12}CH_4$ used in the nominal experiments.

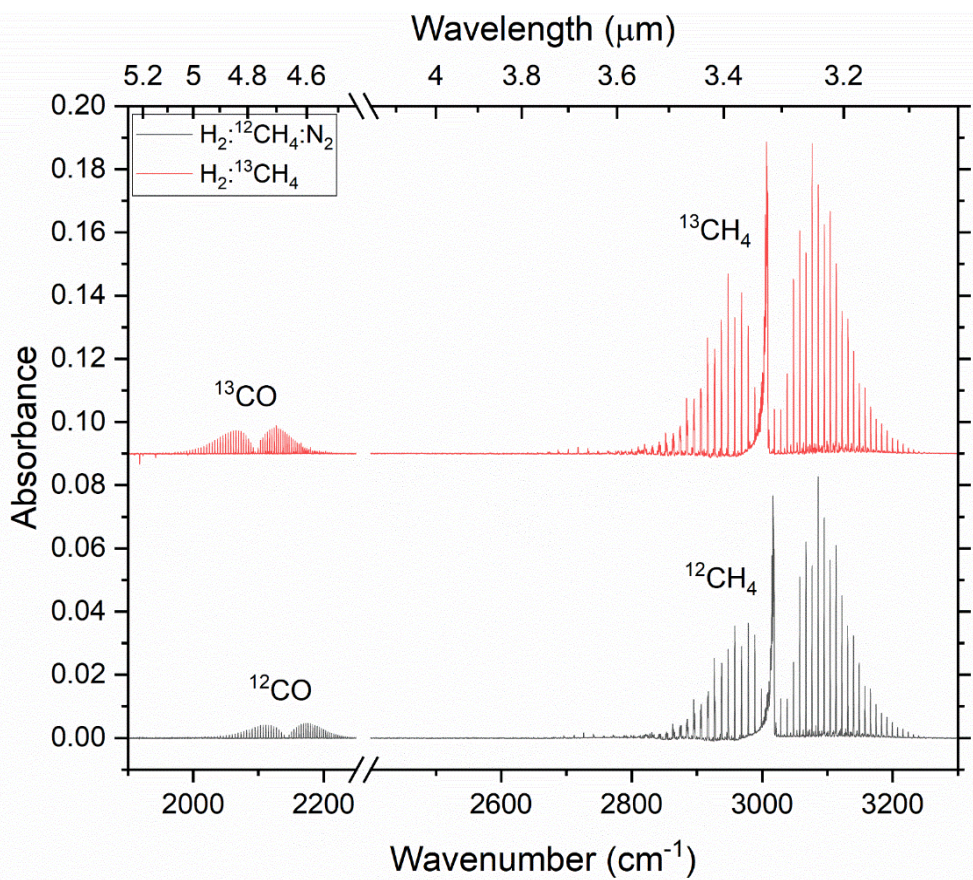

*Figure 5: IR absorbance spectra of the $H_2$:$^{12}CH_4$:$N_2$ (99%:0.5%:0.5%) gas mixture (bottom) and the $H_2$:$^{13}CH_4$ (99.4 %:0.6%) gas mixture (top) after 22 hr of heating at 1073 K. Spectra have been offset for clarity.*

Figure 5 presents the IR absorbance spectra of the $H_2$:$^{12}CH_4$:$N_2$ (99%:0.5%:0.5%) gas mixture (bottom) and the $H_2$:$^{13}CH_4$ (99.4%:0.6%) gas mixture (top) after 22 hr of heating at 1073 K. As a result of the isotopic labeling of $CH_4$ with $^{13}C$ atoms, we observed a shift of the $\nu_3$ C-H stretching band of $CH_4$ for the $H_2$:$^{13}CH_4$ mixture from 3017 cm$^{-1}$ (3.314 µm) to 3009 cm$^{-1}$ (3.017 µm). Similarly, we observed for the CO formed after the heating of the gas a shift of the center of the $\nu_3$ absorption band from 2143 cm$^{-1}$ (4.666 µm) for the $H_2$:$^{12}CH_4$:$N_2$ gas mixture to 2095 cm$^{-1}$ (4.773 µm) for the $H_2$:$^{13}CH_4$ gas mixture. This indicates that the CO formed in these conditions is also labelled with $^{13}C$ atoms and confirms that the source of carbon for the formation of CO is effectively the initial $CH_4$. Our data set is not sufficient to determine in detail the thermochemical mechanism responsible for the formation of CO during our experiments. However, methane oxidation into CO could follow the net reaction R1 (Line, et al. 2010; Madhusudhan et al. 2011; Moses, et al. 2011; Visscher & Moses 2011), which is endothermic and can occur only at high temperature, in agreement with the observation of CO formation only at the highest studied temperatures (873 and 1073 K). Furthermore, this reaction could potentially be catalyzed on the walls of the cell:

$CH_4 + H_2O \rightarrow CO + 3H_2$ (R1)

The amount of CO produced after the heating of both gas mixtures at 873 and 1073 K was quantified using the method described in Section 2.2 and the results are presented in Table 2.

*Table 2: Partial pressures of CO after 24 hr of heating of the $H_2$:$CH_4$:$N_2$ and $H_2$:$CH_4$:$H_2O$ gas mixtures at 873 and 1073 K. The uncertainties are given at 2 standard deviations and were calculated from the standard fluctuations of the infrared spectroscopy measurements*

|  | $H_2$:$CH_4$:$N_2$ | $H_2$:$CH_4$:$H_2O$ |
|---|---|---|
| $T$ (K) | $P_{CO}$ (µbar) | $P_{CO}$ (µbar) |
| 873 | 1.3 ± 0.2 | 1.4 ± 0.1 |
| 1073 | 8.0 ± 0.2 | 8.3 ± 0.2 |

We obtained similar results for the experiments conducted with the $H_2$:$CH_4$:$H_2O$ gas mixture and the corresponding IR spectra are presented in Figure 6. Depending on the experiments, we observed negative or positive variations of the $CO_2$ and $H_2O$ absorbance. This can be explained by variations, during the experiments, of the air contents ($CO_2$ and $H_2O$) in the optical pathway outside of the cell. While in the case of $H_2$:$CH_4$:$H_2O$ (98.4%:0.8%:0.8%), we would have expected much stronger $H_2O$ absorption bands (based on the absorption cross sections of $H_2O$ compared to the ones of $CH_4$), we do not see this during our experiments. This indicates that, after degassing, the surface of the tube is depleted of a large fraction of adsorbed $H_2O$ and becomes activated. Subsequently, when the gas mixture is introduced at room temperature in the cell, $H_2O$ from the mixture is immediately adsorbed on the cell walls, showing far less or no gas-phase $H_2O$ in the spectra shown in Figure 6. This possibility also explains why $H_2O$ absorption fluctuates, which is mainly due to the atmospheric $H_2O$ fluctuations outside the cell in the path of the IR beam. Further support for this possibility comes from the fact that we do not observe any significant difference in the amount of CO produced by thermochemistry whether the initial gas mixture contained water or not (see Table 2 above) – indicating the source of oxygen for the thermal oxidation chemistry is likely $H_2O$ molecules adsorbed on the interior wall surface at the cold extremities of the cell or an extremely small amount of air leaking into the cell during the experiments. For this reason, we could not track the evolution of the water amount in the cell during the heating of the gas. We initially expected to present the results of the experiments with and without $H_2O$, for comparison, but it appears that assessing the role of $H_2O$ on the chemistry would require a different experimental approach, to better control the amount of water present in the gas phase during the different phases of the experiments, as well as complete mitigation of the adsorption of water on the colder regions of the cell wall. Therefore, in the rest of this article, we will only present the results of the experiments with $N_2$.

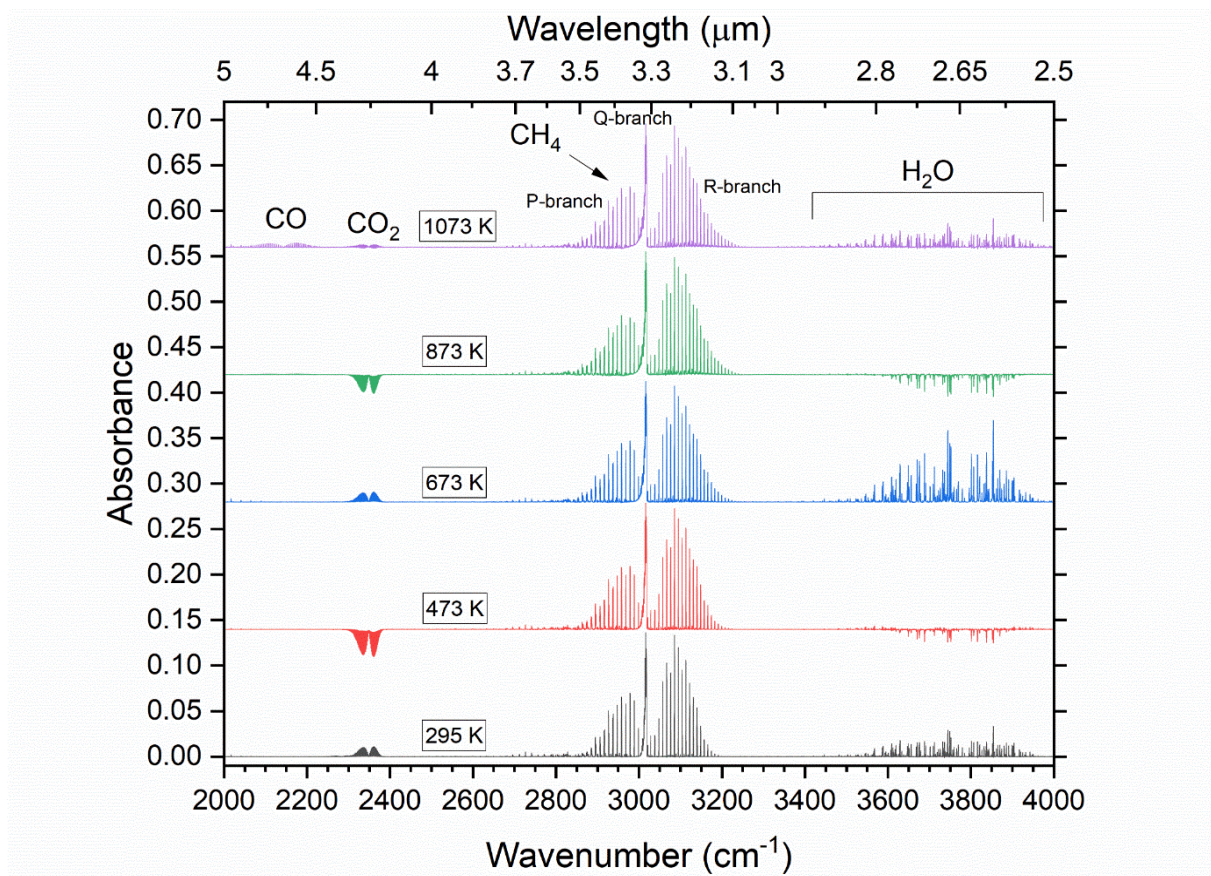

*Figure 6: IR absorbance spectra of the initial gas mixture $H_2$:$CH_4$:$H_2O$ (98.4%:0.8%:0.8%) at ambient temperature (black) and after 22 hr of heating at different set temperatures: 473 K (red), 673 K (blue), 873 K (green), and 1073 K (violet). Spectra have been offset for clarity.*

### 3.2. Experimental Gas-phase Photochemistry

#### 3.2.1. $CH_4$ Consumption

The first step of the gas-phase photochemistry is the consumption of the initial reactants $H_2$, $CH_4$, and $N_2$. Among them, $CH_4$ is the only molecule that can be directly photodissociated by the photons emitted by our UV lamp and its evolution is monitored using IR spectroscopy. To better visualize changes after irradiation, we calculated the difference of absorbance between spectra after irradiation ($A_{irradiation}$) and spectra before irradiation ($A_{heating}$) presented in Figure 4. The resulting spectra are presented in Figure 7, which focuses on the 2600-3500 cm$^{-1}$ (2.857 - 3.846 µm) range. Absorption bands that appear with negative values correspond to species that have been consumed during the irradiation, while bands with positive values correspond to absorption bands of species that have been produced during the irradiation. In Figure 7, we observed for all studied temperatures a decrease of the methane absorbance in the P, Q, and R branches, indicating a consumption of $CH_4$ during the irradiation.

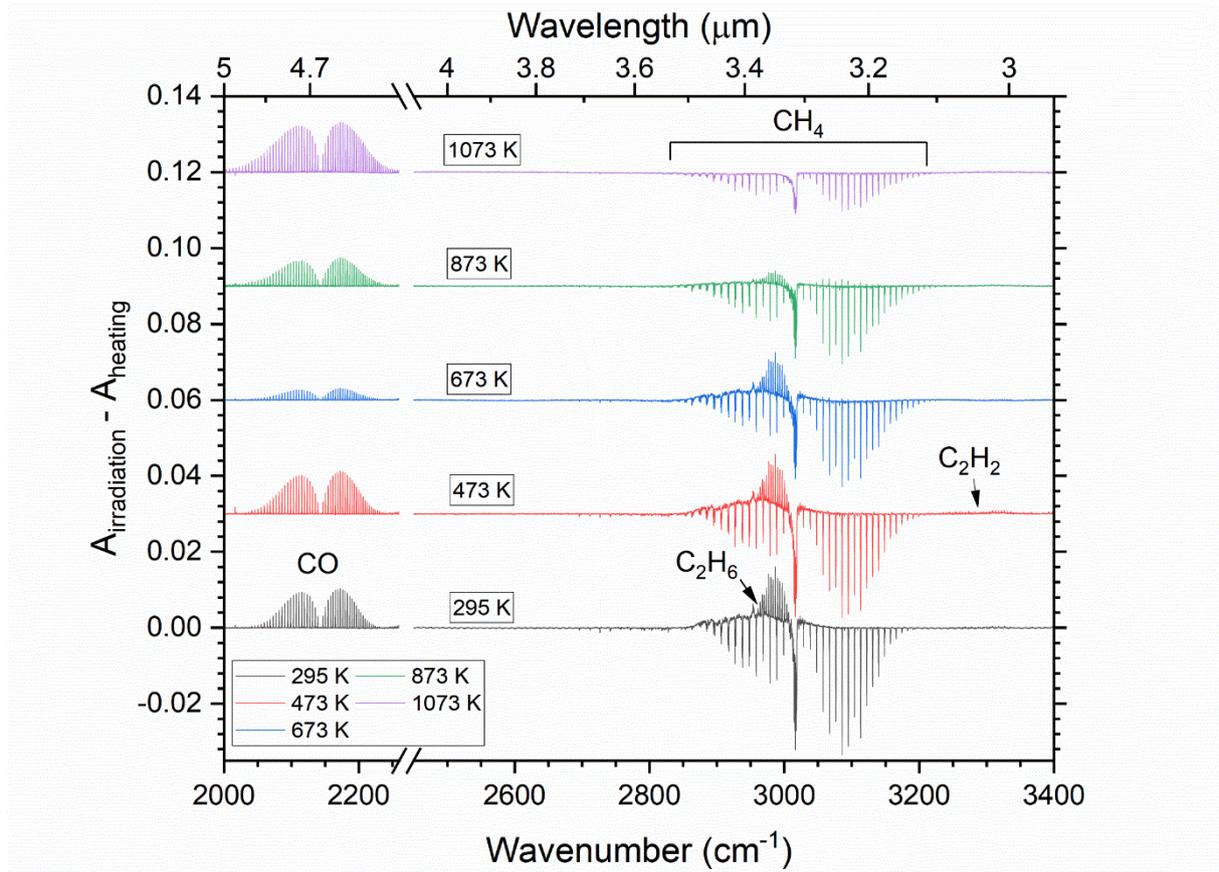

*Figure 7: Difference of absorbance after irradiation ($A_{irradiation}$) and before irradiation ($A_{heating}$) for the $H_2$:$CH_4$:$N_2$ (99%:0.5%:0.5%) gas mixture and for different gas temperatures: ambient temperature (black) and different set temperatures: 473 K (red), 673 K (blue), 873 K (green), and 1073 K (violet). Spectra have been offset for clarity.*

Following the procedure described in Section 2.2, we determined from the different IR spectra the partial pressure of $CH_4$ (in µbar) for all studied conditions (temperatures and gas compositions). We decided to quantify $CH_4$ using its R branch only, to avoid possible overlapping with absorption bands of other species (i.e., $C_2H_6$) that could result in higher uncertainties. The results are presented in Table 3.

*Table 3: Methane partial pressures before $P_0(CH_4)$ and after $P_t(CH_4)$ irradiation for all the studied temperatures for the $H_2$:$CH_4$:$N_2$ gas mixture as determined from IR spectra as well as methane consumption efficiency $e_{CH_4}$. The uncertainties are given at 2 standard deviations and were calculated from the standard fluctuations of the infrared spectroscopy measurements. A graphical comparison of the evolution of the $e_{CH_4}$ and of the main product formation efficiencies as a function of the temperature is shown in Figure 10.*

| $T$ (K) | $P_0$ ($CH_4$) (µbar) | $P_t$ ($CH_4$) (µbar) | $e_{CH_4}$ (%) |
|---|---|---|---|
| 295 | 72 ± 2 | 43 ± 1 | 40 ± 2 |
| 473 | 96 ± 3 | 63 ± 2 | 35 ± 2 |
| 673 | 115 ± 3 | 85 ± 2 | 26 ± 1 |
| 873 | 126 ± 4 | 98 ± 3 | 22 ± 1 |
| 1073 | 132 ± 4 | 111 ± 3 | 16 ± 1 |

An example of comparison between the measured and simulated absorbances (following the procedure described in Section 2.2) for the quantification of CH$_4$ in the case of the H$_2$:CH$_4$:N$_2$ gas mixture after 22 hr at 295 K is shown in Figure 8. In general, we observed a good agreement between the position and intensities of the calculated and the measured spectra. However, we observed some discrepancies in the intensities of the rotational transitions for the higher rotational levels. This could be the result of uncertainties in the temperature profile being used to calculate the gas spectrum, which may vary compared to the real temperature of the gas.

For a quantitative comparison of the methane consumption as a function of the experimental conditions, we have calculated the methane consumption efficiency for each experiment. We defined the CH$_4$ consumption efficiency $e_{CH_4}$ (in percent) according to the following equation (2):

$$e_{CH_4} = \frac{P_0(CH_4) - P_t(CH_4)}{P_0(CH_4)} \tag{2}$$

where $P_0(CH_4)$ and $P_t(CH_4)$ are the methane partial pressures (in µbar) before irradiation and after a time *t* of irradiation, respectively.

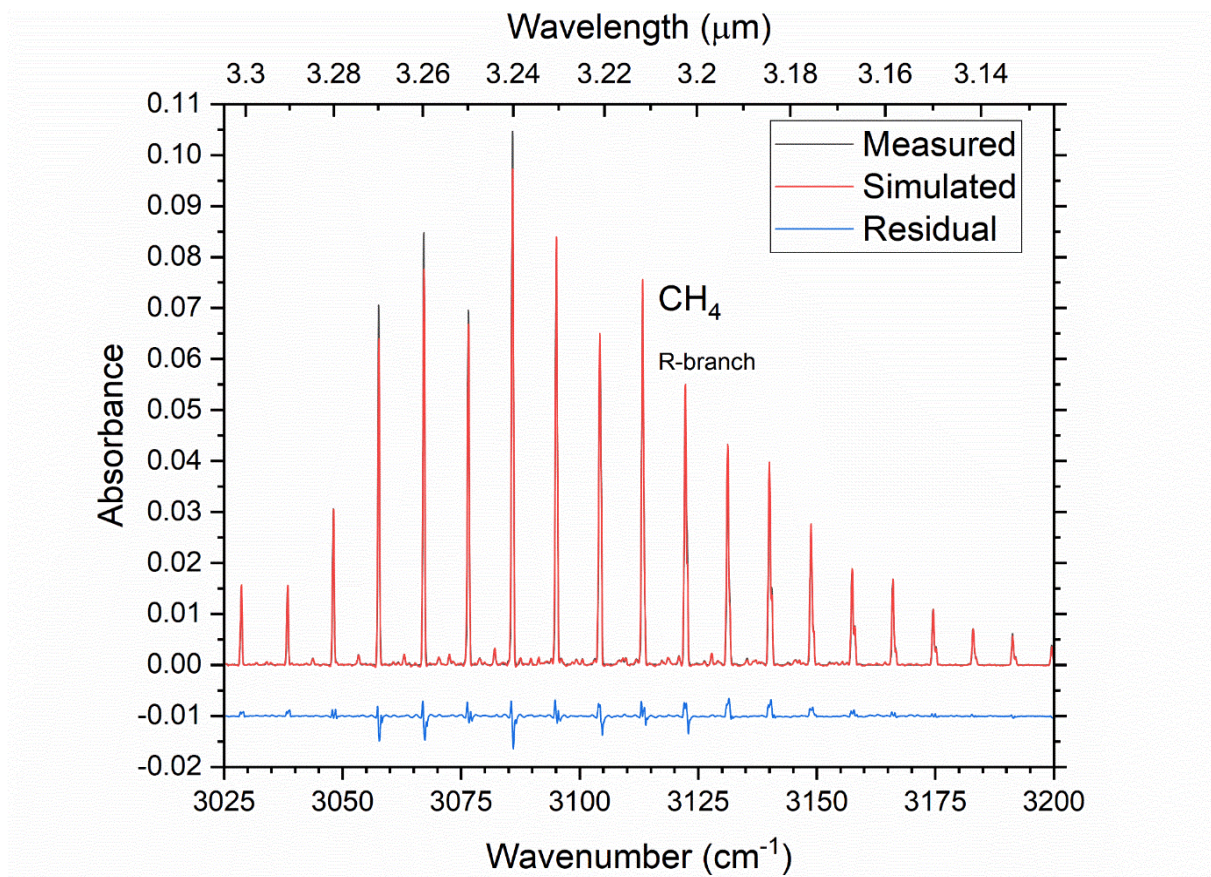

*Figure 8: Comparison of measured and simulated absorbances of $CH_4$ for a spectral resolution of 0.45 cm$^{-1}$ in the case of the $H_2$:$CH_4$:$N_2$ gas mixture after 22 hr at 673 K. The simulated spectrum was obtained following the procedure described in Section 2.2. Residual (measured - simulated) is shifted vertically for clarity.*

We observed a decrease of the methane consumption efficiency when the temperature of the gas increases: the consumption efficiency decreases from 40% at ambient temperature to ~16% at 1073 K. More experimental data will be needed to fully understand the chemistry of $CH_4$ in our experiment and explain the decrease of the methane consumption efficiency with increase in temperature. However, the more likely explanation is an increase of the rates of reactions that recycle methane's photoproducts ($CH_3$, $CH_2$, etc.) back into $CH_4$.

### 3.2.2. Formation of Photochemical Products

#### 3.2.2.1. Hydrocarbons

We used IR spectroscopy to identify gaseous products formed during the experiment. In the differential IR spectra presented in Figure 7, we observed two systems of absorption bands with positive values, indicating that they can be attributed to new species formed during the irradiation.

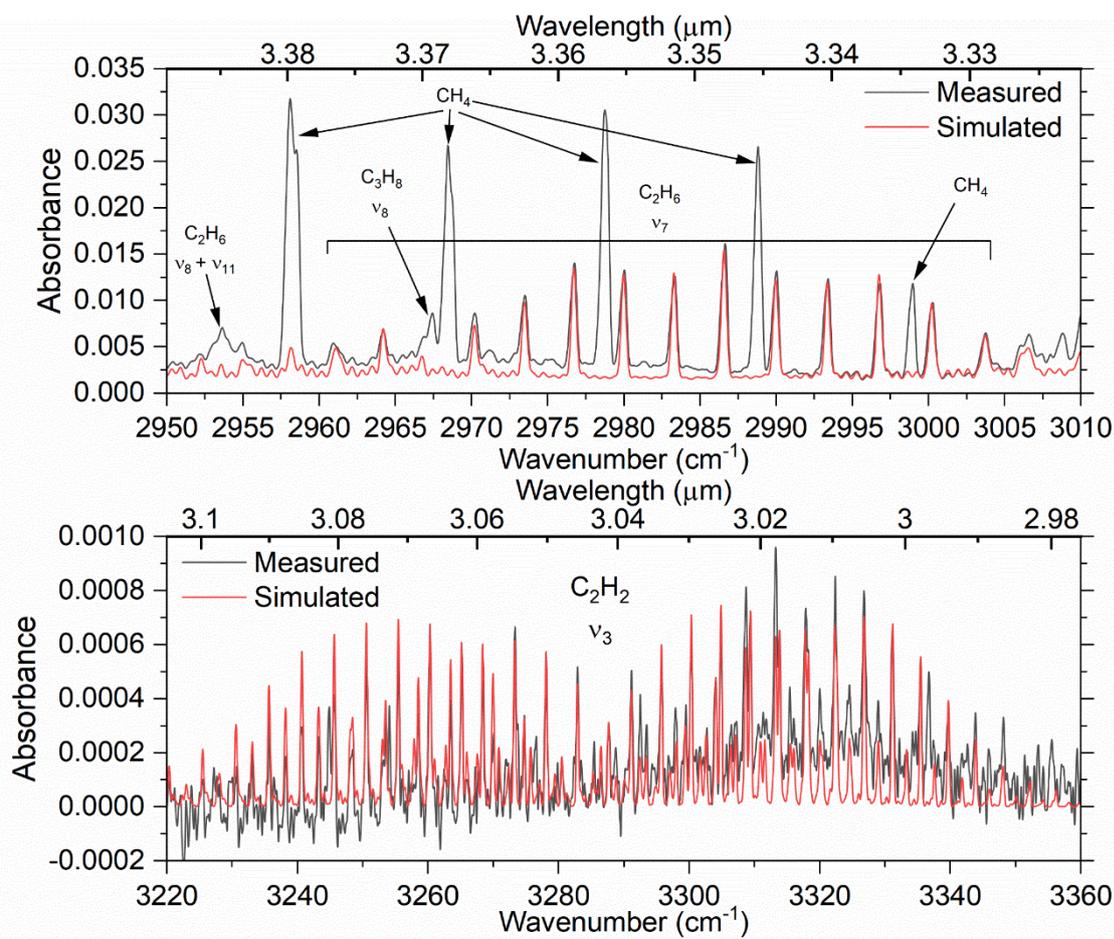

*Figure 9: Comparison of measured and simulated absorbances of $C_2H_6$ (top) and $C_2H_2$ (bottom) for a spectral resolution of 0.45 $cm^{-1}$ in the case of the $H_2$:$CH_4$:$N_2$ gas mixture after 24 hr of irradiation at 473 K. The simulated spectra were obtained following the procedure described in Section 2.2.*

The first band system covers the 3200-3350 $cm^{-1}$ (2.985 - 3.125 µm) range and is centered at ~ 3285 $cm^{-1}$ (3.044 µm). This can be unambiguously attributed to the rotational transitions of the P and R branches of the $v_3$ C-H asymmetric stretching band of $C_2H_2$ (Vanderauwera et al. 1993). The second band system, which overlaps with the P branch of the $CH_4$ absorption band, covers the 2959-3030 $cm^{-1}$ (3.3 - 3.379 µm) range and is centered at ~2987 $cm^{-1}$ (3.347 µm). Those bands can be attributed unambiguously to the rotational transitions of the Q branch of the $v_7$ stretching bands of $C_2H_6$ (Hargreaves et al. 2015; Harrison et al. 2010; Lattanzi et al. 2011) while rotational transitions of the P and R branches are not visible in our spectra. In addition to the $v_7$ stretching bands, we observed a weak absorption band at ~2954 $cm^{-1}$ (3.385 µm), which can also be attributed to $C_2H_6$ and corresponds to the combination band $v_8 + v_{11}$ (Lattanzi, et al. 2011). Figure 9 presents an example of comparison between measured and simulated (following the procedure described in Section 2.2) absorbances after the quantification process of $C_2H_6$ and $C_2H_2$, in the case of the $H_2$:$CH_4$:$N_2$ gas mixture after 24 hr of irradiation at 473 K. For $C_2H_6$, we observed a good agreement between the position and intensities of the calculated and the measured spectra for the $\nu_7$ absorption band. However, the $\nu_8 + \nu_{11}$ combination band is not present in the line list we used to calculate the absorbance spectrum of $C_2H_6$ (Gordon, et al. 2022), therefore no comparison could be done for this band. Finally, the position of the rovibrational band system for $C_2H_2$ is in good agreement with the simulated spectra (following the procedure described in Section 2.2), confirming its formation. However, the intensities of these bands do not show good matching between experiment and simulation. It should be noted that the absorbance of $C_2H_2$ is about 300 times weaker than the $CH_4$ band and close to the noise floor. As a result, the low signal-to-noise ratio for this band in our experimental spectrum is the limiting factor for the quantification of $C_2H_2$ production. For other experimental conditions, the intensity of this band system was too low to unambiguously confirm the presence of the $C_2H_2$ in the spectrum or to quantify it. Finally, when comparing the experimental spectra with a synthetic calculated spectrum of $C_2H_6$ as presented in Figure 9 (top), we observed an absorption band at 2967 cm$^{-1}$ (3.37 µm) that is convoluted with a rotational transition of the R branch of the $\nu_3$ stretching band of $CH_4$. This absorption does not fall into the rotational transition of the $\nu_7$ stretching band of $C_2H_6$ but is shifted to higher frequency by a few wavenumbers. We assign this band to the $\nu_8$ stretching band of propane, $C_3H_8$, (Harrison & Bernath 2010), indicating the formation of higher hydrocarbons during the UV photolysis.

The $C_2H_2$, $C_2H_6$, and $C_3H_8$ formation in our experiments can only be explained by chemical pathways involving radicals formed by the dissociation of $CH_4$. Indeed, the energy of the photons emitted by our vacuum UV (VUV) lamp and transmitted though the $MgF_2$ window ($\lambda > 115$ nm) is not sufficient to photoionize the molecules present in our initial gas mixture and initiate chemical pathways involving ion-molecule reactions. Methane photodissociation and branching ratios for the different channels have been determined with precision by Gans et al. (2011) at Ly$_\alpha$, the dominant wavelength responsible for the methane photolysis in our experiments. At this wavelength, methane photodissociation follows the four pathways below with their corresponding branching ratios (Gans, et al. 2011). $CH_n(X)$ corresponds to radicals in their fundamental states, while $CH_n(a)$ corresponds to radicals in their first excited state:

$CH_4 + h\nu \rightarrow CH_3(X) + H$          0.42          (R2a)

$\rightarrow CH_2(a) + H_2$          0.48          (R2b)

$\rightarrow CH(X) + H_2 + H$          0.07          (R2c)

$\quad \rightarrow CH_2(X) + H + H \quad\quad 0.03$ (R2d)

Following $CH_4$ photodissociation, $C_2H_6$ can be formed by the following termolecular reaction (R3) involving the methyl radical $CH_3(X)$, which is one of the two main products of the $CH_4$ photolysis:

$CH_3 + CH_3 + M \rightarrow C_2H_6 + M$ (R3)

In addition, we determined, thanks to the analysis of our 0D thermo-photochemical model (see Section 2.3) that in our $H_2$-dominated gas mixture, $CH_2(a)$ could also lead to the $CH_3$ radicals, through either reaction with $H_2$, or collisional deexcitation to $CH_2(X)$ followed by reaction with H atoms promoting the formation of $C_2H_6$ in agreement with our experimental results.

Then, because of its strong absorption cross section at $Ly_\alpha$ (Chen & Wu 2004), the $C_2H_6$ formed in the cell can be photolyzed following these five different pathways with associated branching ratios (Akimoto et al. 1965):

$C_2H_6 + h\nu \rightarrow C_2H_4 + H_2 \quad\quad 0.14$ (R4a)

$\quad \rightarrow C_2H_2 + H_2 + H_2 \quad\quad 0.27$ (R4b)

$\quad \rightarrow CH_3 + CH_3 \quad\quad 0.06$ (R4c)

$\quad \rightarrow CH_4 + CH_2(a) \quad\quad 0.22$ (R4d)

$\quad \rightarrow C_2H_4 + 2H \quad\quad 0.31$ (R4e)

Hence, the $C_2H_2$ observed in our experiment can be formed directly from the photodissociation of $C_2H_6$ with the reaction R4b or indirectly from the successive photodissociation of $C_2H_4$, which is the main product of $C_2H_6$ photodissociation by the reactions R4a and R4e. In that case, $C_2H_2$ is produced by one of the two following reactions (Balko et al. 1992; Lee et al. 2004; Vuitton et al. 2019):

$C_2H_4 + h\nu \rightarrow C_2H_2 + H_2 \quad\quad 0.5$ (R5a)

$\quad \rightarrow C_2H_2 + 2H \quad\quad 0.5$ (R5b)

Competitively, $C_2H_4$ molecules can react with H radicals in the following termolecular reaction to produce $C_2H_5$ radicals (Dobrijevic et al. 2016).

$C_2H_4 + H + M \rightarrow C_2H_5 + M$ (R6)

Then $C_2H_5$ radicals can react with H radicals in a termolecular reaction to reform $CH_3$ radicals (R7) or they can react with $CH_3$ radicals in another termolecular reaction (R8) to produce $C_3H_8$ molecules (Dobrijevic, et al. 2016), which is the last hydrocarbon product detected in our experiments:

$C_2H_5 + H + M \rightarrow CH_3 + CH_3 + M$ (R7)

$C_2H_5 + CH_3 + M \rightarrow C_3H_8 + M$ (R8)

Other and more complex hydrocarbons may have been formed during our experiments, but in too low abundance to be detected by our IR spectrometer. Indeed, as it can be seen with the case of $C_2H_2$ in Figure 9, absorption bands with an absorbance lower than $1 \times 10^{-3}$ have a low signal-to-noise ratio and absorption bands with an absorbance lower than a few $10^{-4}$ would likely be not detected. For a species with an absorption cross section like the one of $C_2H_2$, the limit of detection would correspond to a partial pressure lower than 1 µbar as $C_2H_2$ has been detected with a partial pressure of 1.5 µbar (see Table 4 below).

From the IR spectra, we have quantified the partial pressures of $C_2H_6$ and $C_2H_2$ after irradiation, following the method described in Section 2.2. For $C_3H_8$, because the band is blended with a rotational transition of $CH_4$, we could not quantify it. The results are provided in Table 4.

*Table 4: Partial pressures of $C_2H_6$, CO, and $C_2H_2$ after 24 hr of irradiation of the $H_2$:$CH_4$:$N_2$ gas mixture at the different studied temperatures. The uncertainties are given at 2 standard deviations and were calculated from the standard fluctuations of the infrared spectroscopy measurements.*

| T (K) | $P_{C2H6}$ (µbar) | $P_{CO}$ (µbar) | $P_{C2H2}$ (µbar) |
|---|---|---|---|
| 295 | 11 ± 1 | 9.3 ± 1.0 | - |
| 473 | 15 ± 2 | 14 ± 2 | 1.5 ± 0.2 |
| 673 | 17 ± 2 | 5 ± 1.0 | |
| 873 | 8.0 ± 1.0 | 13.5 ± 0.5 | - |
| 1073 | - | 47 ± 3 | - |

After the analysis of the potential chemical pathways occurring in our experiment, we evaluated the relative efficiency of the carbon conversion from methane to the different hydrocarbon molecules whose abundances have been quantified (i.e., $C_2H_2$ and $C_2H_6$) for the different studied temperatures. This relative efficiency (in percent) has been calculated for each temperature by dividing the partial pressure of the species determined from the IR spectra after irradiation (Table 4) by the initial partial pressure of $CH_4$ measured before irradiation (Table 3). The results are presented in Figure 10 along with the methane consumption efficiencies (Table 3). First, we observed that at every studied temperature, $C_2H_6$ is the most abundant hydrocarbon product and most of the consumed methane reacted to form $C_2H_6$. In addition, we observed that the amount of ethane produced decreases when the temperature increases, in correlation with the decrease of the methane consumption efficiency. In a modeling study of the influence of temperature and atmospheric composition on atmospheric photochemistry, Adams et al. (2022) found that when the temperature increases from less than 180 K to ~500 K in an $H_2$-dominated atmosphere, reaction of $CH_3$ radicals with $H_2$ would recycle back $CH_3$ to methane (R9) at a faster rate than the termolecular reaction involving two $CH_3$ radicals to form $C_2H_6$ given above (R3).

$CH_3 + H_2 \rightarrow CH_4 + H$ (R9)

Such a mechanism agrees with our experimental findings, although other mechanisms, including the reaction of $CH_3$ radicals with OH due to the presence of water (see Section 3.2.2.2) could also be considered. However, it should also be noted that despite the agreement with our experimental results, these mechanisms do not appear to match the theoretical simulations that we performed in this study (see Section 3.3 below). Additional studies will be necessary to fully understand how the temperature affects the chemistry in our experiments and confirm or reject this hypothesis.

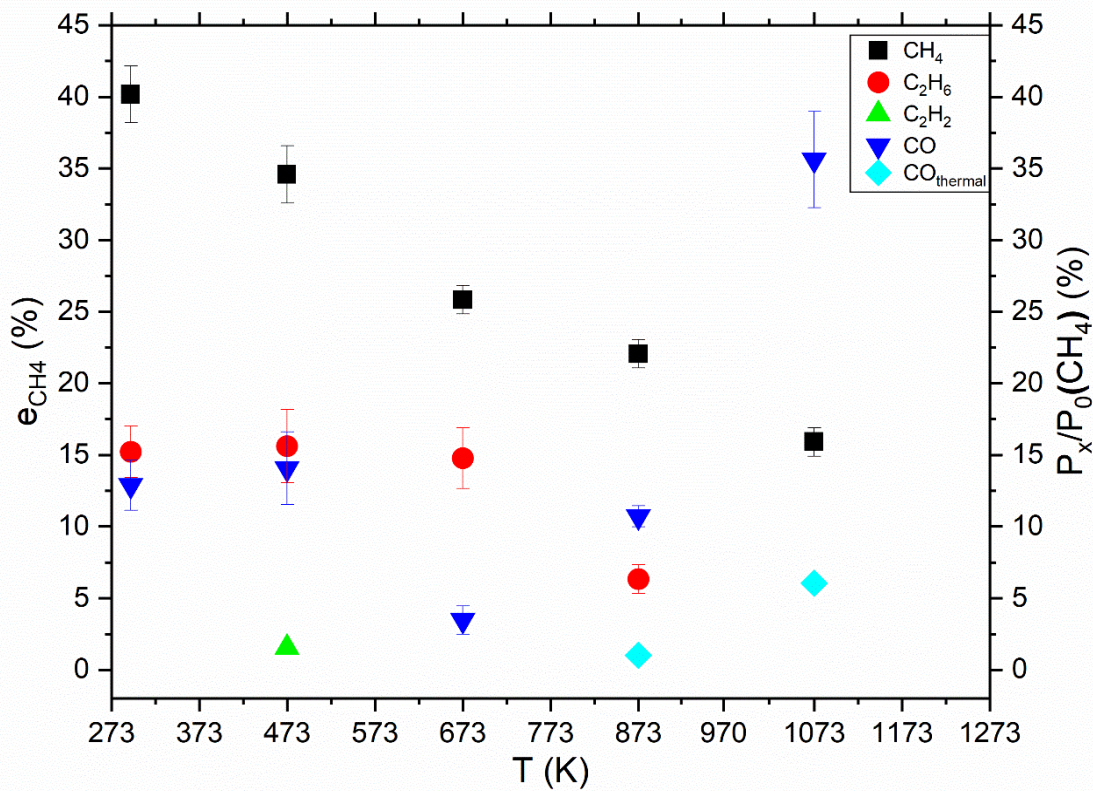

*Figure 10: Mixing ratios of CO after 22 hr of heating, methane consumption efficiency ($e_{CH_4}$), and mixing ratios of $C_2H_6$, CO, and $C_2H_2$ after 24 hr of irradiation of the $H_2:CH_4:N_2$ gaseous mixtures at ambient temperature (295 K) and the different set oven temperatures 473, 673, 873, and 1073 K.*

#### 3.2.2.2. Carbon Monoxide

In addition to the formation of hydrocarbons, Figure 7 shows the formation of CO after irradiation, identified by its absorption band at 2143 cm$^{-1}$, at every studied temperature, while thermochemical formation of CO was observed only at the highest studied temperatures (Section 3.1). As for other species, we have quantified the amount of CO produced in every condition using the method described in Section 2.2. The results are presented in Table 4 and an example of comparison between the measured spectrum and the one calculated during the quantification process is presented in Figure 11 in the case of the $H_2:CH_4:N_2$ gas mixture after irradiation at 295 K. When comparing the two spectra, we observe a good agreement in the band positions, but some discrepancies in the intensity of some of the rotational transitions, which result in an increase of the uncertainties on the CO quantification. This could be due to a saturation of the most intense rovibrational transitions. As for hydrocarbons, we calculated the relative efficiency of the $CH_4$ to CO conversion for the different studied conditions and the results are shown in Figure 10. At 295 K and 473 K, the amounts of CO and $C_2H_6$ produced are similar, given the error bars. The CO production at 673 K is significantly lower than $C_2H_6$, but at 873 and 1073 K, the production of CO is higher than the one of $C_2H_6$, and it is correlated with the decrease of the $C_2H_6$ production. CO becomes the most abundant product of the $CH_4$ photochemistry at these two temperatures.

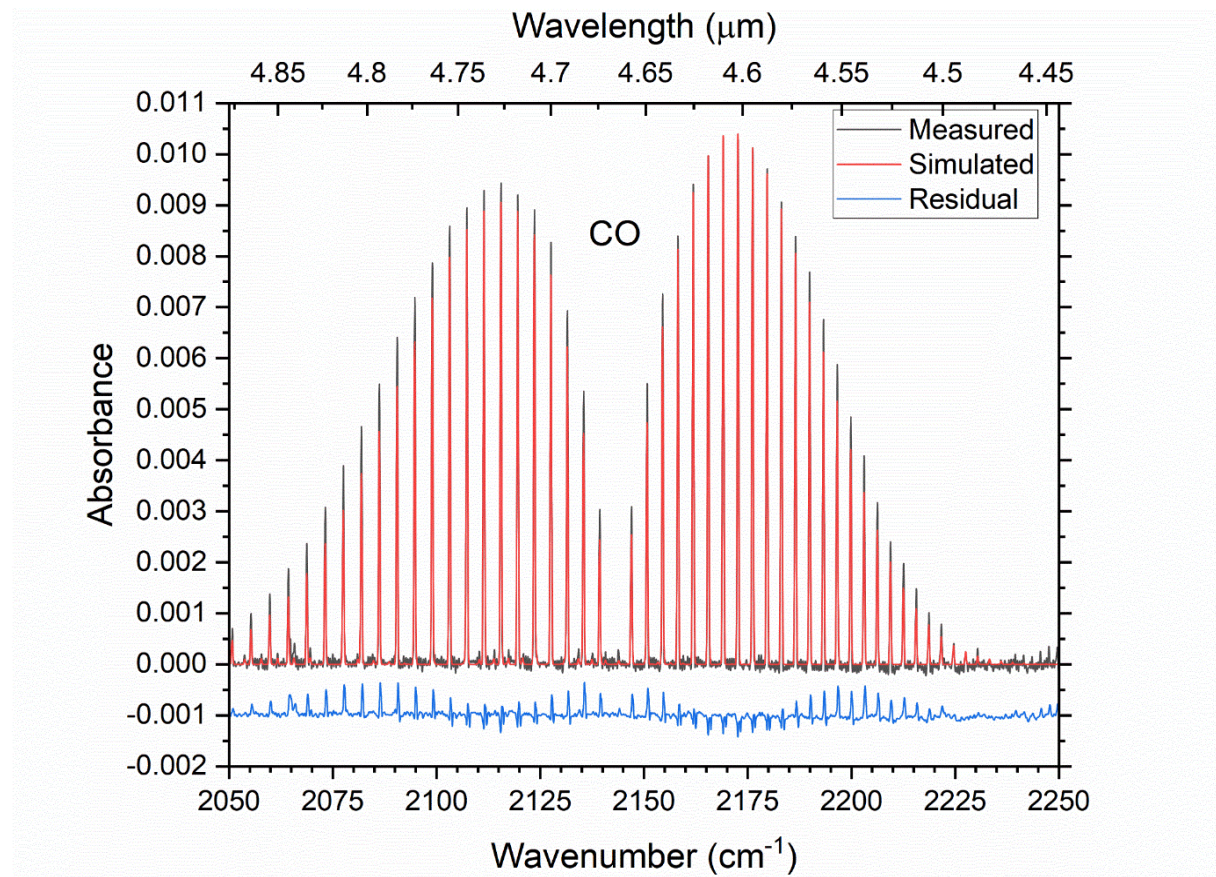

*Figure 11: Comparison of measured and simulated absorbances of CO for a spectral resolution of 0.45 cm$^{-1}$ in the case of the $H_2$:$CH_4$:$N_2$ gas mixture after 24 hr of irradiation at 295 K. The simulated spectrum was obtained following the procedure described in Section 2.2. Residual (measured - simulated) is shifted vertically for clarity.*

Although the formation of a small amount of CO was observed at 873 and 1073 K after the heating of the gaseous mixtures, we observed here that photochemistry drastically enhanced the formation of CO at every studied temperature, as shown in Figure 10. The higher formation rate of CO in the presence of VUV photons could be explained by the photodissociation of water desorbing from the cell walls, as discussed in Section 3.1 and in Fleury, et al. (2019), and/or collisional reactions of photolyzed methane ($CH_3$, $CH_2$, and CH radicals) with the surface of the cell walls. Water photodissociation will principally produce H atoms and hydroxyl radicals (OH) following the reaction R10:

$H_2O + h\nu \rightarrow OH + H$ (R10)

OH radicals can then participate in the conversion of $CH_4$ into CO via the formation of methanol ($CH_3OH$) or $CH_2OH$ radicals, which are important intermediates of the net reaction R1 (Line, et al. 2010; Madhusudhan & Seager 2011; Moses, et al. 2011; Visscher & Moses 2011), although none of these species were detected in our experiments. $CH_3OH$ has absorption bands around 2941 cm$^{-1}$ (3.4 µm) (Harrison et al. 2012), which is within the spectral range covered by our IR spectrometer. Given the limit of detection of our spectrometer, as discussed in Section 3.2.2.1, these bands would have needed to be a few $10^{-4}$ or more in absorbance to be detected. This corresponds to a partial pressure of 1 µbar or more, depending on the absorption cross sections of the molecules compared to $C_2H_2$. $CH_3OH$ and $CH_2OH$ can be notably formed by the reaction of hydroxyl radicals with methyl radicals following the reactions R11 and R12, respectively (Jasper et al. 2007; Visscher & Moses 2011).

$CH_3 + OH + M \rightarrow CH_3OH + M$ (R11)

$CH_3 + OH \rightarrow CH_2OH + H$ (R12)

Thereby, water photodissociation could have enhanced the conversion of $CH_4$ to CO and at the same time have inhibited the ethane production. Indeed, in this case, the reactions of OH with $CH_3$ (R11 and R12) would compete with the termolecular reactions between two $CH_3$ to form $C_2H_6$ (R3), decreasing its production efficiency, in agreement with our experimental results.

### 3.2.3. Nitrogen Chemistry

In addition to hydrocarbons, the formation by disequilibrium chemistry (quenching and photochemistry) of nitrogen-bearing species, such as ammonia ($NH_3$) and hydrogen cyanide (HCN), is predicted by different theoretical studies using thermo-photochemical models (Moses, et al. 2011; Venot, et al. 2015; Venot, et al. 2012; Venot et al. 2013b). Therefore, it has been proposed that the detection of these species in the atmospheres of certain exoplanets could also be an indicator of the effect of disequilibrium chemistry in these atmospheres (MacDonald & Madhusudhan 2017). However, we did not observe in this study the formation of N-bearing compounds for the experiments made with the $H_2:CH_4:N_2$ gas mixture. As mentioned in Section 2.1, this is likely explained by the fact that in our experiments, the gas mixture is irradiated by photons with λ > 115 nm, with more energetic photons being absorbed by the $MgF_2$ window closing our reaction cell. Hence, $N_2$ cannot be photodissociated or photoionized, and thus nitrogen chemistry was not initiated in our experiments.

### 3.3. Simulations of the Experiments with a 0D Thermo-photochemical Model

We have simulated with a 0D thermo-photochemical model our laboratory experiments to compare the chemical composition observed in our experiments with the one predicted by a kinetic model adapted to high temperatures, with the same initial conditions (e.g., pressure, temperature, gas composition, etc.).

First, we have calculated with the model the evolution of the methane consumption efficiency $e_{CH_4}$ (in percent) as a function of time for the different studied temperatures: 295, 473, 673, 873, and 1073 K. The results are presented in Figure 12.

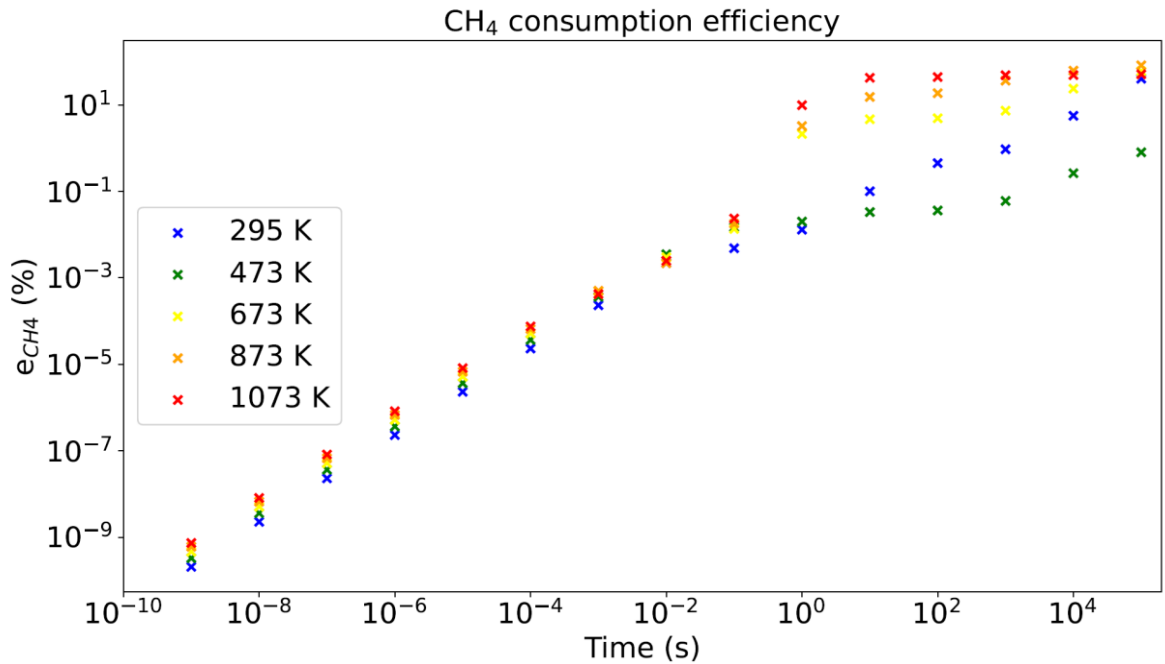

*Figure 12: Evolution of the methane consumption efficiency ($e_{CH_4}$) as a function of time of irradiation obtained with a 0D thermo-photochemical model. Simulations were made for an $H_2$:$CH_4$:$N_2$ (99%:0.5%:0.5%) initial gas mixture, a pressure of 15 mbar, and different gas temperatures: 295, 473, 673, 873, and 1073 K. $e_{CH_4}$ at $10^5$ s are presented as function of the temperature in Figure 13a.*

At every temperature, $e_{CH_4}$ first increases rapidly as a function of time until $10^{-1}$ s, rising by 8 orders of magnitude. Then $e_{CH_4}$ continues to increase, but at a rate depending on the temperature until $10^5$ s. As set in the initial conditions (see Section 2.3), the consumption efficiency, calculated by the model after 24 hr (40%) at 295 K is similar to the one observed in our experiment (i.e., 40%). When increasing the temperature of the gas, first at 473 K, we observed that the model predicts a decrease of the methane consumption efficiency (lower than 1%) compared to the value predicted for the ambient temperature. Then, at higher gas temperatures, the predicted value of $e_{CH_4}$ increases and becomes higher than the value at 295 K with respectively, 65% at 673 K and 80% at 873 K. Finally, the methane consumption efficiency decreases again at 1073 K with a value of ~50%, which remains higher than the value obtained at ambient temperature. Except for the value obtained at 473 K, we observed with our 0D kinetic model that the methane consumption efficiency is enhanced by the increase of the temperature. These results are opposite to the ones obtained experimentally, where the $e_{CH_4}$ decreased from 40 to 16% as the temperature increased from 295 to 1073 K. The temporal evolution observed at 473 K does not seem to be consistent with that observed for the other temperatures. Such surprising behavior will be the subject of a future study, to determine if it is a real phenomenon expected at this temperature or a consequence of a bad representation of specific reaction rates. We note that the model used is developed and validated for high-temperature applications (Venot, et al. 2020a; Venot, et al. 2012). The two lowest temperatures studied here are in the lower limit of the application domain and therefore subject to more uncertainty. More studies are needed to understand why the global trend obtained with our model (with the higher temperatures) is in contradiction with the experimental results. Second, we have calculated from the model's results at $10^5$ s the relative efficiency (see Section 3.2.2.1) of the carbon conversion from methane to the hydrocarbon molecules whose abundances have been quantified in the experiments (i.e., $C_2H_2$ and $C_2H_6$) for the different studied temperatures. This relative efficiency (in percent) has been calculated for each temperature by dividing the abundances of these species ($n_x$) at $10^5$ s by the initial amount of $CH_4$ ($n_0(CH_4)$). The results are presented in Figure 13 (top) along with the methane consumption efficiency at $10^5$ s (Figure 12). First, we observed that the $C_2$ hydrocarbon production rates predicted by the model are an order of magnitude lower than the one measured experimentally. Second, the model predicts that at every studied temperature, $C_2H_2$ is two to 10 times more abundant than $C_2H_6$, which is consistent with what is observed in modeling studies of warm gas giant atmospheres (Line, et al. 2011; Moses 2014; Venot, et al. 2015). However, as for the methane consumption efficiency, this is in contradiction with what has been observed in our experiments (see Figure 10) in which $C_2H_6$ is more abundant than $C_2H_2$. Moreover, we do not observe a drop of hydrocarbon production efficiency with temperature in our model, but rather an increase in hydrocarbon production efficiency, contrary to what has been observed experimentally. These contradictions are discussed in the next section.

As discussed previously, the unexpected formation of CO in our experiments implies that a source of oxygen is present in our reaction cell, most likely $H_2O$. To evaluate the impact of $H_2O$ on the chemistry and determine if this could explain the observed differences in the production of hydrocarbons between the experimental and the modelling results, we have run a second series of simulations in which we added 0.01% of $H_2O$ to the initial gas mixture composition. As for the simulations without $H_2O$, we have calculated from the model's results at $10^5$ s the relative efficiency (in percent) of the production of $C_2H_2$, $C_2H_6$, as well as CO. Results are presented in Figure 13 (bottom) along with the methane consumption efficiency at $10^5$ s.

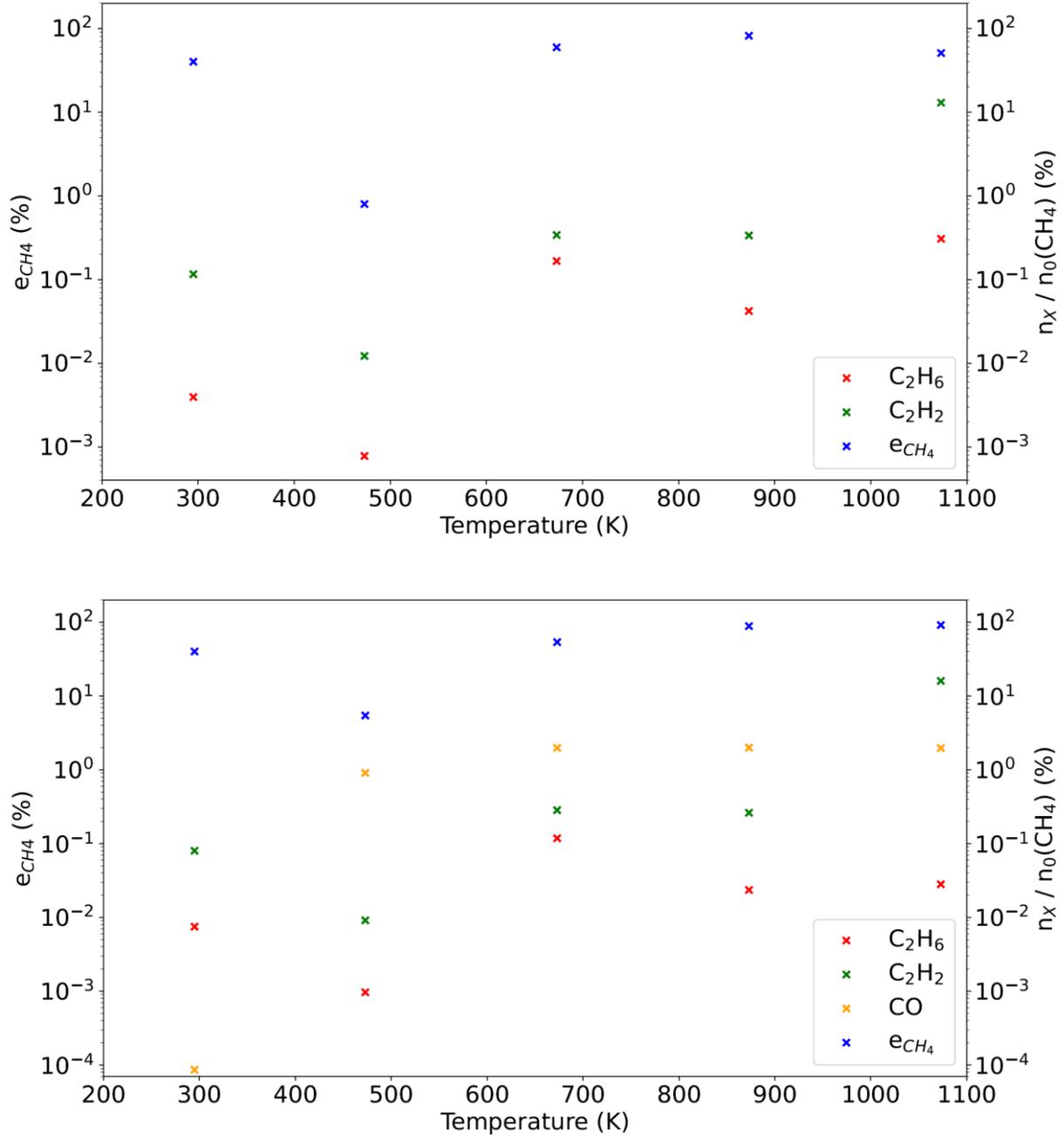

*Figure 13: Top: methane consumption efficiency ($e_{CH_4}$) and mixing ratios of $C_2H_6$ and $C_2H_2$ after $10^5$ s as calculated by the 0D thermo-photochemical model for an $H_2:CH_4:N_2$ (99%:0.5%:0.5%) gas mixture at different temperatures: 295, 473, 673, 873, and 1073 K. Bottom: methane consumption efficiency ($e_{CH_4}$) and mixing ratios of $C_2H_6$ and $C_2H_2$ after $10^5$ s as calculated by the 0D thermo-photochemical model for an $H_2:CH_4:N_2:H_2O$*

*(98.99%:0.5%:0.5%:0.01%) gas mixture at different temperatures: 295, 473, 673, 873, and 1073 K*

We observed in Figure 13 (bottom) that the inclusion of $H_2O$ leads to the production of CO, which remains low at ambient temperature ~$10^{-4}$% and then increases significantly to ~1% at 473 K and above. These simulations confirm that in our experimental conditions, the inclusion of $H_2O$ can result in the production of CO as observed experimentally. However, at every temperature, the production rates of CO calculated in the simulations are drastically lower than the ones obtained experimentally. Moreover, the addition of $H_2O$ also affects, but to a smaller extent, the consumption efficiency of $CH_4$. In particular, $e_{CH_4}$ increases at 473 K compared to the value obtained in the simulations without $H_2O$. Finally, the addition of $H_2O$ also affects the production of $C_2H_2$ and $C_2H_6$, which increase or decrease depending on the temperature. However, regardless of whether the inclusion of $H_2O$ allows the simulation to reproduce qualitatively the production of CO that we observed experimentally, it does not resolve the contradictions between the model's predictions and the experimental results, as described above.

## 4. Discussion: Comparison with the Hydrocarbon Chemistry in Thermo-photochemical Models for Warm Exoplanet Atmospheres

In our laboratory experiments, we observed that the methane photochemistry led to an efficient production of hydrocarbons, dominated by the production of $C_2H_6$. On the contrary, the simulations we performed with a 0D thermo-photochemical model, as well as other studies of warm gas giant exoplanet atmospheres with thermo-photochemical models, predict that $CH_4$ photochemistry mostly leads to the production of $C_2H_2$, followed by $C_2H_4$ and $C_2H_6$ (Line, et al. 2011; Moses 2014; Venot, et al. 2015). Therefore, differences exist between the chemical pathways leading to the production of the $C_2$ hydrocarbons in kinetic models and in our experiments.

In our experimental conditions, the chemical processes are initiated with the photolysis of $CH_4$, which produces mostly $CH_3$ and $CH_2$ radicals (Gans, et al. 2011). Since $C_2H_6$ was the main hydrocarbon product in our experiments, one can expect that the dominant reaction should be the termolecular reaction involving two $CH_3$ radicals (R3) that produces ethane. However, the $C_2H_6$ that is formed should in turn be photodissociated resulting in the production of $C_2H_2$ and $C_2H_4$ (R4a, R4b, and R4e). Since the formation reaction of ethane (R3) is slower than the ethane photodissociation (R4), the initial reservoir of $C_2H_6$ should progressively be converted into $C_2H_2$ and $C_2H_4$ that become more abundant. However, this is in contradiction with the fact that $C_2H_6$ remains the major product observed in our experiments, implying that some of these reactions may differ in our experimental conditions. Our current experimental data are too limited to firmly explain the origin of this apparent contradiction in the concentration of $C_2H_6$ compared to $C_2H_2$ and $C_2H_4$. Nevertheless, the higher $C_2H_6$ abundance experimentally observed suggests some hypotheses. First, the reaction rate of R3 could increase if it were catalyzed by the walls of the cell, leading to an enhanced production of ethane. Second, the high ethane abundance could be the result of a lower photodissociation rate of $C_2H_6$ into $C_2H_2$ and $C_2H_4$. Third, it could also indicate the existence of competitive mechanisms that efficiently recycle $C_2H_6$, such as the hydrogenation of $C_2H_2$ and $C_2H_4$ by reaction with atomic or molecular hydrogen. Indeed, a modeling study of Titan-like exoplanet atmospheres found that hydrogenation reactions of $C_2$ hydrocarbons such as $C_2H_2$ are favored in $H_2$-dominated atmospheres compared to $H_2$-poor ones (Adams, et al. 2022).

It should be also noted that some differences between our experiment and the model that we used may be responsible for the differing results. First, for this study, we used a 0D model that does not consider the existence of a gradient of temperature along the cell, nor the fact that the actinic flux is lower at the center of the cell, where the temperature is maximal. If in our experiments the chemistry is different in the colder part of the cell than in the warmer part, this effect would not be considered by the model. Combined with the fact that in our experiments, the IR diagnostic probes the gas composition with the contributions of both the cold and warm gases, this would mean that we could minimize experimentally the effect of the chemistry at high temperature compared to what is predicted by the model. Second, we cannot discard the possibility that wall reactions affect the chemistry in our experiments. The 0D model does not consider wall effects, which could also bias the comparison between laboratory and modeling works. Similar conclusions are drawn in a recent reaction-mechanism-generation modeling work by (Yang et al. 2023), which could qualitatively reproduce some of the results of the experiments we performed in Fleury, et al. (2019, 2020), but could not reproduce them quantitatively.

Definitively solving this problem would require performing a kinetic monitoring of the chemical composition of the gas phase, which is beyond our current experimental capabilities. This would allow us to track the formation of the different species and monitor the evolution of their relative ratios as a function of the duration of the irradiation. In addition, detecting and quantifying the formation of intermediate species, including radicals, would help to identify key reactions and build a chemical network for our experiments that could be compared with the ones issued by thermo-photochemical models to identify key differences in the simulated chemistry. Moreover, the relevance of comparison between experimental and modeling results could be improved by developing a more accurate model of our experiments, such as a 1D kinetic model that would consider the existence of a gradient of temperature.

**Conclusion**

We used the CAAPSE experimental setup to study the temperature dependence of the formation of hydrocarbons in warm gas giant atmospheres with $T < 1073$ K. Complementarily, we have run a series of numerical simulations using a 0D thermo-photochemical model with parameters (gas composition, temperature, etc.) mimicking our experimental conditions, to help interpret the results of the experiments.

First, we observed experimentally that in the studied conditions thermochemistry had a negligible impact on the gas-phase composition, except for the highest studied temperatures (i.e., 873 and 1073 K) for which the production of a small amount of CO was observed.

On the contrary, we observed that UV photochemistry drastically affects the gas-phase composition, driving it away from the thermochemical equilibrium. The formation of various hydrocarbons, including $C_2H_6$, $C_2H_2$, and $C_3H_8$ as well as CO, was observed.

We also find that the abundances of the chemical species formed in the experiments are temperature-dependent. Indeed, the increase of the gas temperature induces a decrease of the methane consumption efficiency, as well as an inhibition of the hydrocarbons products' formation at temperatures higher than 673 K, with no hydrocarbon detected at 1073 K. Further experimental work, including kinetic studies of reactants' consumption and products' formation, will be necessary to explain the observed changes in the hydrocarbons' production efficiency with temperature.

Finally, we observed some discrepancies between our experimental and modeling results, highlighted by an important quantitative difference in the production of acetylene and ethane. The fact that ethane was unexpectedly produced in higher quantity than acetylene in our laboratory experiments demonstrates the importance of experimental studies together with modeling studies to accurately interpret, understand, and predict observations of exoplanet atmospheres.

## Acknowledgments

The experimental part of this research was carried out at the Jet Propulsion Laboratory, California Institute of Technology, under a contract with the National Aeronautics and Space Administration. B.F., M.S.G., and B.H. acknowledge support from the NASA Exoplanet Research Program. B.F. thanks the Université Paris-Est Créteil (UPEC) for funding support (postdoctoral grant). O.V. acknowledges funding from the ANR project 'EXACT' (ANR-21-CE49-0008-01), from the Centre National d'Etudes Spatiales (CNES), and from the CNRS/INSU Programme National de Planétologie (PNP). © 2023 California Institute of Technology, all rights reserved.